\documentclass[aps, pra,twocolumn, showpacs,superscriptaddress,10pt,floatfix]{revtex4}
\usepackage[english]{babel}
\usepackage[latin1]{inputenc}
\usepackage{hyperref}
\usepackage{amsmath, amssymb, amsfonts}
\usepackage{amssymb,xcolor}
\usepackage{graphicx}
\usepackage{exscale}
\usepackage{latexsym}
\usepackage{cases}
\usepackage{epstopdf}
\usepackage{subfigure}
\usepackage{bm}
\usepackage{color}
\usepackage{amssymb}

\newcommand{\ket}[1]{|#1\rangle}
\newcommand{\bra}[1]{\langle #1|}
\newcommand{\tr}{\mathrm{Tr}}

\newcommand{\n}{\nonumber\\}

\newcommand{\ex}[1]{\langle #1\rangle}

\allowdisplaybreaks

\begin{document}

\title{Entanglement, Coherence, and Extractable Work in Quantum Batteries}

\author{Hai-Long Shi}
\thanks{These authors contributed equally to this work.}
\address{School of Physics, Northwest University, Xi'an 710127, China}
\address{State Key Laboratory of Magnetic Resonance and Atomic and Molecular Physics, Wuhan Institute of Physics and Mathematics, APM, Chinese Academy of Sciences, Wuhan 430071, China}
\affiliation{University of Chinese Academy of Sciences, Beijing 100049, China}
\author{Shu Ding}
\thanks{These authors contributed equally to this work.}
\address{School of Physics, Northwest University, Xi'an 710127, China}

\author{Qing-Kun Wan}
\address{State Key Laboratory of Magnetic Resonance and Atomic and Molecular Physics, Wuhan Institute of Physics and Mathematics, APM, Chinese Academy of Sciences, Wuhan 430071, China}
\affiliation{University of Chinese Academy of Sciences, Beijing 100049, China}
\author{Xiao-Hui Wang}
\email{xhwang@nwu.edu.cn}
\address{School of Physics, Northwest University, Xi'an 710127, China}
\address{
Shaanxi Key Laboratory for Theoretical Physics Frontiers, Xi'an 710127, China
	}
\address{
Peng Huanwu Center for Fundamental Theory, Xi'an 710127, China
}
\author{Wen-Li Yang}
\email{wlyang@nwu.edu.cn}
\address{
	Institute of Modern Physics, Northwest University, Xi'an 710127, China}
\address{
Shaanxi Key Laboratory for Theoretical Physics Frontiers, Xi'an 710127, China
	}
\address{
Peng Huanwu Center for Fundamental Theory, Xi'an 710127, China
}

\date{\today}

\begin{abstract}
We investigate the connection between quantum resources and extractable work in quantum batteries.
We demonstrate that quantum coherence in the battery or  the battery-charger entanglement  is a necessary  resource for generating nonzero extractable work during the charging process.
At the end of the charging process, we also establish a tight link of coherence and entanglement with the final extractable work: coherence naturally promotes the coherent work  while coherence and entanglement inhibit the incoherent work.
We also show that obtaining maximally coherent work is faster than obtaining maximally incoherent work.
Examples ranging from the central-spin battery and the Tavis-Cummings battery to the spin-chain battery are  given to illustrate these results.

\end{abstract}

\maketitle
\emph{Introduction.}--
The rapid development of modern technology to probe and manipulate physical systems at the qubit level requires a new reconstruction of three laws of thermodynamics by including dominant quantum features \cite{Binder18,Hatsopoulos76,Toyabe10,Baugh05,Alemany10,Brandao15,Muller18,Cwiklinski15,Lostaglio15,Skrzypczyk14,Bera17}.
One of the most serious tasks is to 
identify the impact of quantum resources, such as entanglement \cite{Llobet15,Andolina19,Alicki13,Oppenheim02,Allahverdyan04,Kamin20,Liu21}, coherence \cite{Lostaglio15-2,Uzdin15,Korzekwa16,Caravelli21,Cakmak20,Francica20,Monsel20,Oliviero21}, and discord \cite{Manzano18,Francica17,Francica22}, on extractable work.  
Quantum resource theories \cite{Brandao13,Chitambar19,Streltsov17,Horodecki09,Goold16} provide a possible route to deal with it, but troublesome couplings among different quantum resources that appear in finite-time nonequilibrium thermodynamic processes make the problem complicated and intractable. 
Different quantifiers of extractable work based on  free energy, ergotropy, or  process-dependent work operators  may cause further confusion \cite{Brandao13,Allahverdyan04,Niedenzu19,Aberg14,Skrzypczyk14,Allahverdyan15,Talkner07,Llobet17}.
Recent studies on quantum batteries have contributed greatly to  the investigation of these problems \cite{Andolina19,Kamin20,Liu21,Francica20,Pintos20,Barra19}.

Quantum batteries (QBs) are designed to store energy efficiently for  subsequent use by exploiting various quantum resources.
A paradigmatic model of a quantum battery has been experimentally implemented based on an organic microcavity \cite{Quach22}.
Subject to the second law of thermodynamics, not all energy injected into the battery can be extracted, and the useful part is called extractable work.
Up to now, considerable attention has been mostly focused on identifying useful quantum resources that speed up energy charging dynamics or improve the storage of extractable work. 
Distinguished from individual QB cells, collective QB cells do display a quantum correlation-induced speedup in the energy charging process \cite{Campaioli17,Gyhm22,Ferraro18,Rossini20,Binder15,Le18}.
Recent works focusing on the Tavis-Cummings (TC) battery and the central-spin (CS) battery have illustrated an unfavorable effect of the battery-charger entanglement on the storage of extractable work \cite{Andolina19,Liu21}.
Another effort is to isolate the quantum correlation contributions to  ergotropy \cite{Francica22,Francica20}. 
Despite such progress, it remains vital to establish some general connections between entanglement, coherence, and extractable work.

In this Letter, we shed some light on the clarification of the role of entanglement and coherence on extractable work by providing a model-independent analysis for general QBs.
We present Theorem 1 to emphasize that generating nonzero extractable work requires nonvanishing entanglement or coherence during the charging process.
By restricting to incoherent QBs, whereby only diagonal elements are involved, we prove  an inverse relationship between the final extractable work and the battery-charger entanglement; see Theorem 2.
For a general QB, inspired by Refs. \cite{Francica20,Shi20}, we treat extractable work as a sum of incoherent (extractable) work and coherent (extractable) work.
We show that the former is limited by the diagonal entropy of the battery, i.e., the sum of entanglement and  coherence,
while the latter is promoted by the battery's coherence.
Furthermore, we argue that achieving maximally coherent work  is faster than achieving maximally incoherent work.
We illustrate our results in some concrete models: the CS battery, the TC battery, and the $XXZ$ battery.
These results hold regardless of using free energy work or ergotropy as the quantifier of extractable work.
Our work also deepens the understanding of the role of quantum resources played in energy transfer and work extraction.

\emph{Preliminaries.}--
A general Hamiltonian for describing the charging process can be written as 
\begin{eqnarray}\label{Hamiltonian}
\mathcal H(t)=\mathcal H_b+\mathcal H_c+\mathcal V(t), 
\end{eqnarray}
where the local Hamiltonians $\mathcal H_b$ and $\mathcal H_c$ characterize the battery part and the charger part, respectively.
The charging operator $\mathcal V(t)$ incorporates all terms that control
the switch of the energy injection, such as battery-charger interactions or some external driven fields.
At time $t=0$, the whole system is  prepared in a product state $\rho_{bc}(0)$ with the battery being the ground state $\ket{\varepsilon_0}$ of  $\mathcal H_b$. 
We then suddenly turn on $\mathcal V$ and aim to inject as much energy as possible into the battery for a finite time interval $[0,T]$.
Such a time interval $T$ is called the charging time.  
The evolving state is given by $\rho_{bc}(t)=U(t)\rho_{bc}(0)U^\dag(t)$ with $U(t)=\mathcal T\exp(-i\int_0^t\mathcal H(t) dt)$. 

An important physical quantity is the energy that can be extracted from the battery state $\rho_b(t)=\tr_c(\rho_{bc}(t))$.
The first extracted work quantifier is given by the free-energy difference between $\rho_b(t)$ and a specified thermal state $\tau_\beta=\exp(-\beta\mathcal H_b)/Z$
\begin{eqnarray}\label{W_f}
W_f(t)=F(\rho_b(t))-F(\tau_\beta),
\end{eqnarray}	 
where $\beta$ is the inverse temperature and the free energy is $F(\rho_b)=\tr(\rho_b\mathcal F)$ with $\mathcal F=\mathcal H_b+\beta^{-1}\log\rho_b $.
Such work can be extracted in single-shot scenarios by correlating the battery to catalytic systems \cite{Muller18} or in the limit of asymptotically many copies of the battery \cite{Brandao13}. 
Another one is the ergotropy \cite{Allahverdyan04}  which is defined as the most energy that can be extracted by using an optimal cyclic unitary transformation 
$W_e(t)=\max_{U\in\mathcal U_c}\left[ E(\rho_b(t))-E(U\rho_b(t)U^\dag) \right]
$ 
where $E(\rho_b)=\tr(\rho_b\mathcal H_b)$ and $\mathcal U_c$ denotes the set of all cyclic unitary transformations.
A close expression for this expression can be obtained in terms of the passive state $\tilde \rho_b(t)=\sum_nr_n \ket{\varepsilon_n}\bra{\varepsilon_n}$ 
\begin{eqnarray}\label{W_e}
W_e(t)=E(\rho_b(t))-E(\tilde \rho_b(t)),
\end{eqnarray} 
where the eigenstates of $\mathcal H_b=\sum_n\varepsilon_n\ket{\varepsilon_n}\bra{\varepsilon_n}$ and $\rho_b(t)=\sum_nr_n\ket{r_n}\bra{r_n}$ are reordered so that $r_0\geq r_1\geq r_2\geq\cdots$ and $\varepsilon_0\leq \varepsilon_1\leq\varepsilon_2\leq\cdots$.
This kind of quantum work has been experimentally measured recently in a single-atom heat engine \cite{Horne20} and a spin heat engine \cite{Lindenfels19}.

Quantum coherence and entanglement are two fundamental quantum features that reflect different manifestations of the same principle--the superposition principle.
We will focus on them to explore their connections with extractable work.
Coherence of the battery is characterized by the minimum relative entropy $S(\rho_b(t)||\delta)$ of $\rho_b(t)$ with respect to all incoherent states $\delta\in \mathcal I$ \cite{Baumgratz14}:
\begin{eqnarray}\label{C=S_d-S}
C(t)=\min_{\delta\in\mathcal I}S(\rho_b(t)\| \delta)=S_{\Delta}(t)-S(t),
\end{eqnarray}
where the minimum is obtained by the dephased state $\Delta[\rho_b(t)]=\sum_n\ex{\varepsilon_n|\rho_b(t)|\varepsilon_n}\ket{\varepsilon_n}\bra{\varepsilon_n}$  and $S_\Delta(t)=-\tr[\Delta[\rho_b(t)]\log_2(\Delta[\rho_b(t)])]$ is the diagonal entropy.
For a bipartite pure state $\rho_{bc}(t)$, the Von Neumann entropy of the battery's reduced density matrix $S(t)=-\tr[\rho_b(t)\log_2(\rho_b(t))]$  
characterizes its battery-charger entanglement \cite{Nielsen}.
Therefore, from Eq. (\ref{C=S_d-S}) we see that the diagonal entropy $S_\Delta(t)$ quantifies the sum of entanglement and coherence.
Henceforth, entanglement always refers to the battery-charger entanglement.

\emph{Necessity of nonzero entanglement or coherence.}--
As proven in Ref. \cite{Pintos20}, the rate of extractable work changes is given by 
$d W_f/dt=-i\tr([\rho_{bc}(t),\mathcal F]V),$ 
where $V$ describes interactions between the battery and the charger.
If during the charging process $\rho_{bc}(t)$ keeps a product state without coherence in its battery part, then $[\rho_{bc}(t),\mathcal F]=0$ and thus $W_f(t)=W_f(0)=0$.
Here we always assume that there is no extractable work at $t=0$.
For general extractable work quantifiers, we can find the same result:

\emph{Theorem 1}.--
For a product pure state $\rho_b(0)\otimes\rho_c(0)$ as the initial state,
if no battery's coherence and no battery-charger entanglement appear during the charging process then extractable work is always zero.

The core idea behind this theorem is that quantum dynamics is heavily restricted by entanglement and coherence. 
In an  extreme case (no entanglement and no coherence), our proof \cite{SM} reveals that the battery state $\rho_b(t)$ cannot be evolved into a high-energy state and thus possesses no extractable work.  
This theorem emphasizes the necessity of entanglement and coherence for generating nonzero extractable work.
Meanwhile, it also implies that it is possible to charge a battery only using entanglement or only using coherence.
The following examples (the CS battery without coherence and the $XXZ$ battery) will confirm this possibility.
In this sense, entanglement and coherence should be placed on an equal footing to treat.

So how do these  resources specifically determine the behavior of extractable work?
Firstly, we will focus on incoherent QBs to uncover the role of  entanglement.

\emph{Incoherent quantum batteries.}--
Because of the vanishing  coherence, the evolution of the density matrix of the incoherent battery can be parametrized as
$
\rho_b(t)=\sum_n p_n(t) \ket{\varepsilon_n}\bra{\varepsilon_n},
$
where $p_n(t)$ denotes the population of the $n$-th energy level.
The charging process is characterized by exciting  low-energy states to high-energy states.
At the end of the charging process, it is thus expected that the population of low-energy levels is not greater than  that of high-energy levels, i.e., $p_n(T)\leq p_{n+1}(T)$, for an ``excellent'' QB.
Under this ideal assumption, we find  that the final extractable work is enslaved to the battery-charger entanglement in incoherent QBs.

\emph{Theorem 2}.--
The final extractable work (whether defined by $W_e(T)$ or $W_f(T)$) is negatively related to the entanglement $S(T)$ for all incoherent QBs.

We now use Fig. \ref{fig1}(a) to explain the essence of the 
 rigorous proof given in  the Supplemental Material \cite{SM}.
Figure \ref{fig1}(a) plots two energy-level configurations for charged battery states.
Compared with the left configuration, the right configuration has a larger population at high-energy levels. 
This difference will lead to $S(T)>S'(T)$ but $W_{e/f}(T)<W_{e/f}'(T)$, which proves the theorem.

This theorem indicates that vanishing  entanglement  at the end of the charging process is helpful to improve the final extractable work.
It seems to contradict the previous theorem, Theorem 1.
In fact, although the final entanglement is harmful, it is still necessary during the charging process  to generate excited states.
The continuing excitation from low-energy states to high-energy states strengthens the battery-charger entanglement until the population of the battery part displays a nearly balanced distribution.
Further excitation will induce the population inversion and lead to a decrease in entanglement. 
Therefore, at the end of the charging process, the vanishing entanglement characterizes that the battery is located in a high-energy state and thus possesses optimal extractable work.

\begin{figure}[t]
\includegraphics[width=3in]{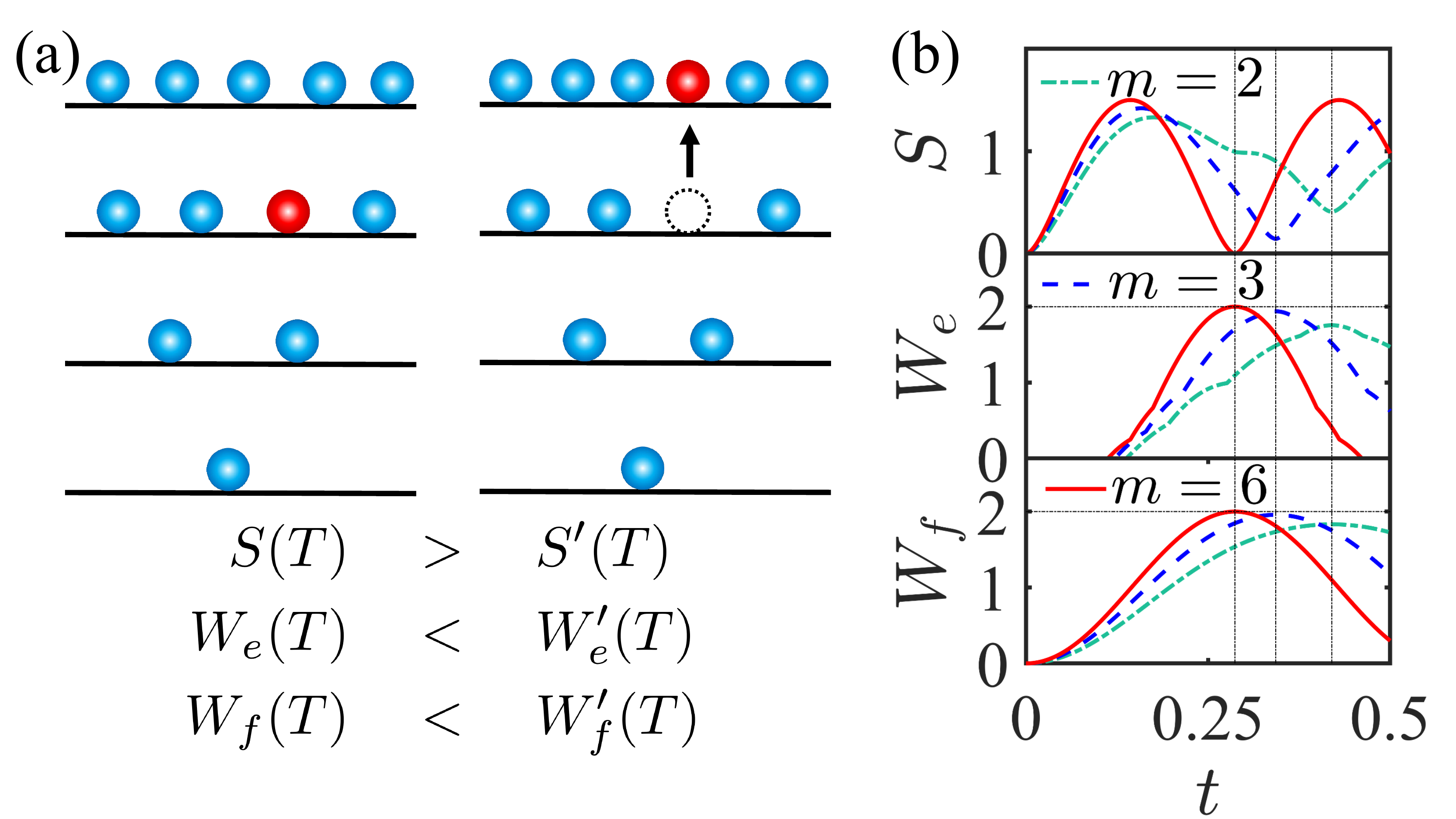}
\caption{
(a) Schematic illustration of the negative relation between  entanglement $S$ and extractable work $W_{e/f}$.
(b) Time evolution of extractable work and entanglement in the CS battery with different Dicke states $\ket{m}$ as initial chargers.
The thin dashed vertical lines from the left to right denote the final charging time $T$ for $m=6,3,2$.
The other parameters are set to $\omega=A=1$, $\beta=100$, $N_b=2$, and $N_c=10$.
}\label{fig1}
\end{figure}

\emph{Example 1: Central-spin battery.}--
The CS quantum battery consists of $N_b$ battery cells and $N_c$ charging units, which are described by collective spin operators	$S^\alpha=\sum_{i=1}^{N_b}\sigma_i^\alpha/2$ and $J^\alpha=\sum_{i=1}^{N_c}\sigma_i^\alpha/2$ $(\alpha=x,y,z)$, respectively.
The charging Hamiltonian reads as
\begin{eqnarray}
\mathcal H_{\rm CS}=\omega S^z+\omega J^z+A(S^+ J^-+S^-J^+),\nonumber
\end{eqnarray}
where $A$ characterizes the flip-flop interaction and $S^\pm=S^x\pm iS^y$.
We prepare  the battery in the ground state $\ket{\downarrow\downarrow\cdots\downarrow}_b$ of $H_b=\omega S^z$ ($\omega>0$) and the charger in a Dicke state $\ket{m}_c$ with $m$ spin-ups.
Because of the flip-flop interaction, the battery will be excited to high-energy states at the cost of decreasing the number of spin-up charging units.
An analytical treatment for this model can be found in Ref. \cite{Liu21}.
In Fig. \ref{fig1}(b), we plot the evolutions of entanglement and  extractable work for three chargers, namely, $m=2,3,6$.
We see that from $m=2$ to $m=6$ the final extractable work $W_{e/f}(T)$ increases, while the final entanglement $S(T)$ decreases,
which is what our Theorem 2 stated. 
Moreover, the almost zero $S(T)$ for $m=6$ indicates  almost optimal work, i.e., $W_{e/f}(T)\simeq W_{e/f}(\ket{\uparrow\uparrow})=2\omega$. 

\emph{General quantum batteries.}--
To emphasize the role of  coherence, it is instructive to isolate the incoherent contribution to the ergotropy, namely, incoherent work  \cite{Francica20}, 
\begin{eqnarray}\label{W_e^i}
W_e^i(t)=E(\Delta[\rho_b(t)])-E(\widetilde{\Delta[\rho_b(t)]}),
\end{eqnarray}
 which is defined as the energy difference between the dephased state and its passive counterpart.
Coherent  work is naturally given by
\begin{eqnarray}\label{W_e^c}
W_e^c(t)&=&W_e(t)-W_e^i(t)\n 
&=&E(\widetilde{\Delta[\rho_b(t)]})-E(\tilde\rho_b(t)).
\end{eqnarray}
An important bound uncovering the role of  coherence is given by \cite{Francica20}
\begin{eqnarray}\label{W_e^c-C}
| W_e^c(t)-\beta^{-1}C(t)|\leq 
\beta^{-1}D(\tilde\rho_b(t)\|\tau_\beta),
\end{eqnarray} 
where $C(t)$ is the battery's coherence defined in Eq. (\ref{C=S_d-S}), $\tilde\rho_b(t)$ is the passive state, $\tau_\beta$ is an arbitrary thermal state with the inverse temperature $\beta$, and $D(\rho\|\sigma)=\tr\{\rho(\log\rho-\log\sigma)\}$ is the quantum relative entropy.
If the passive state can be rewritten as a thermal state, then coherent  work is exactly the  coherence up to a coefficient $\beta^{-1}$.
This condition naturally holds for QBs with only two energy levels,
but it is not satisfied for general QBs.
Although it has been shown that $W_e^c$ is not a coherence monotone in general \cite{Francica20}, we still find that the battery's coherence can qualitatively describe the change of coherent work. 
Figures \ref{fig2}(c) and \ref{fig2}(d) show that coherent work $W_e^c(t)$ is positively related to coherence $C(t)$ 
 in the TC  battery and the $XXZ$ battery.

To unify the different extractable work quantifiers, we take the same strategy to define the incoherent work for the free energy:
\begin{eqnarray}\label{W_f^i}
W_f^i(t)=F(\Delta[\rho_b(t)])-F(\tau_\beta).
\end{eqnarray}
Corresponding coherent work is given by 
\begin{eqnarray}
W_f^c(t)&=&W_f(t)-W_f^i(t)\n
&=&F(\rho_b(t))-F(\Delta[\rho_b(t)]).
\end{eqnarray}
By noticing that  $\rho_b(t)$ and $\Delta[\rho_b(t)]$ have the same energy,
we immediately see an exact correspondence between  free-energy-based coherent work $W_f^c(t)$ and the battery's coherence $C(t)$
\begin{eqnarray}\label{W_f^c-C}
W_f^c(t)=\beta^{-1}C(t).
\end{eqnarray}

Equations (\ref{W_e^c-C}) and (\ref{W_f^c-C}) indicate that quantum coherence can improve coherent work.
So what quantum resource is responsible for incoherent work?

From Eqs. (\ref{W_e^i}) and (\ref{W_f^i}), we see that incoherent  work can be understood as the extractable work from the dephased state $\Delta[\rho_b(t)]$.
Therefore, Theorem 2 is applicable to analyze incoherent work by replacing the von Neumann entropy $S(T)$ with the diagonal entropy $S_\Delta(T)$. 
Explicitly, the smaller the diagonal entropy $S_\Delta(T)$, the larger the incoherent work $W_{e/f}^i(T)$.
Recall from what we discussed earlier that diagonal entropy equals entanglement plus coherence.
So coherence is now   detrimental to incoherent work.
The distinct effects of coherence on coherent work and incoherent work explain why it is so difficult to clarify the role of coherence in extractable work.
The negative relation between incoherent work and the diagonal entropy also reveals that different quantum features will couple to each other and determine the performance of QBs together.

To obtain optimally incoherent work, it is thus expected that the  battery after charging has vanishing coherence and vanishing entanglement, i.e., $S_\Delta(T)=0$.
Thus, incoherent quantum batteries are 
promising candidates, e.g, the CS battery. 
We also observe from the definitions of extractable work [Eqs. (\ref{W_f}) and (\ref{W_e})] that optimal  work  can be achieved when $\rho_b(T)$ is located in the highest energy state, which also requires that  $S_\Delta(T)=0$.
So we identify optimally incoherent  work with optimal work.
On the other hand, to obtain optimally coherent  work, it is naturally expected that $\rho_b(T)$ is in a maximally coherent state.
Because of the complementarity between maximal coherence and entanglement \cite{Singh15}, we give priority to QBs that are charged by only some external fields since no chargers ensures no entanglement.
Next we will give concrete examples to achieve optimally coherent and incoherent work.

\emph{Example 2: TC  battery and XXZ  battery}.--
 The charging Hamiltonian of the TC  battery  reads as
\begin{eqnarray}
\mathcal H_{\rm TC}=\omega S^z+\omega a^\dag a+g(S^+a+S^-a^\dag). \nonumber
\end{eqnarray}
Different from the CS  battery, the TC  battery uses photons as the charger, which is characterized by the creation and annihilation operators $a^\dag$ and $a$.
Performance of the TC battery heavily relies on the probability distribution of the optical field initial state in the number states since the whole Hilbert space can be split into a direct sum of invariant subspace, each containing a component of the initial state $\ket{\psi}_{bc}=\sum_{n=0}^\infty\alpha_n \ket{-N_b/2}_b\ket{n}_c$.
Here the quantum states of the battery part and the charger part are expressed in terms of the Dicke states and the Fock states, respectively.
We introduce the probability amplitudes $\alpha_n^C=e^{-N_c/2}N_c^{n/2}/\sqrt{n!}$ and $\alpha_n^F=\delta_{n,N_c}$ to denote the coherent state and the Fock state with $N_c$ photons, respectively. 
The analytic form for the time evolution operator within the invariant subspaces is  given in Ref. \cite{Zhang18,Delmonte21}.

\begin{figure}[t]
\includegraphics[width=3in]{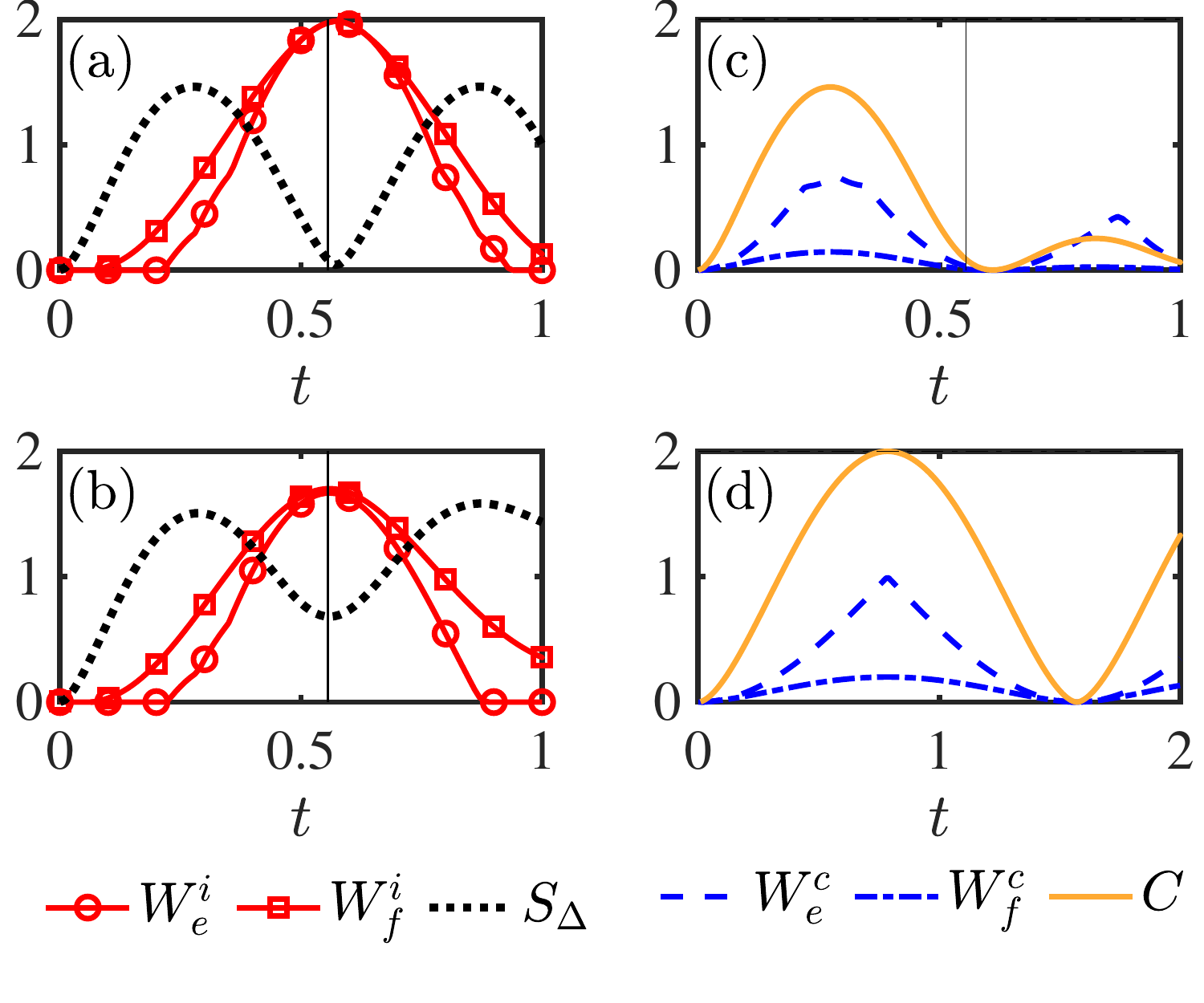}
\caption{
Time evolution of incoherent work, diagonal entropy, coherent  work, and coherence in (a) the TC  battery with an eight-photon Fock state $\ket{8_F}$ as the charger, (b),(c) the TC  battery with an eight-photon coherent state $\ket{8_C}$ as the charger, and (d) the $XXZ$  battery.
The other parameters are set to $\omega=g=1$ for the TC  battery and $\omega=\Omega=J=\Delta=1$ for the $XXZ$  battery.
The inverse temperature is set to $\beta=10$.
}\label{fig2}
\end{figure}

For a two-cell  TC battery ($N_b=2$), time evolution of the population of the highest energy state $\ket{1}_b$ is given by
\begin{eqnarray}
&&\rho_{11}(t)=\sum_{n=2}^\infty \alpha_{n}^2\left[\frac{\sqrt{b_{n}b_{n-1}}}{b_{n}+b_{n-1}}-\frac{\sqrt{b_{n}b_{n-1}}}{b_{n}+b_{n-1}}\cos(\Delta_{n}t)\right]^2\nonumber
\end{eqnarray}
where $b_n=2n$ and $\Delta_n=g\sqrt{b_n+b_{n-1}}$.
It immediately follows that the optimal  work, i.e., $W_{e/f}(T)\simeq 2\omega$ or $\rho_{11}(T)=1$,  can be achieved when we use the Fock state as the charger and assume the number of photons $N_c\gg 1$.
We see from Fig. \ref{fig2}(a) that this result holds even for $N_c=8$ and  the optimal work corresponds exactly to  the disappearance of diagonal entropy.
Different from the Fock state case, however, coherent  work originating from the off-diagonal elements will exist in the coherent state case; see Fig. \ref{fig2}(c).
Although coherence vanishes after the charging process, the optimal  work still cannot be achieved due to the nonvanishing diagonal entropy, see Figs. \ref{fig2}(b) and \ref{fig2}(c).

Figure \ref{fig2}(c) also shows that the battery's coherence reaches its maximum but is not maximally coherent.
To achieve the optimally coherent  work, we consider the $XXZ$ quantum battery \cite{Kamin20}.
The corresponding Hamiltonian reads as
\begin{eqnarray}
&&\mathcal H_{\rm XXZ}(t)=\mathcal H_b+\mathcal V(t),\n
&&\mathcal H_b=\frac{\omega}{2}\sum_{j=1,2}\sigma_j^z,\n
&&\mathcal V(t)=J(\sigma_1^x\sigma_2^x+\sigma_1^y\sigma_2^y+\Delta\sigma_1^z\sigma_2^z)\n
& &+\Omega\sum_{j=1,2}[\cos(\omega t)\sigma_j^x+\sin(\omega t)\sigma^y_j].\nonumber
\end{eqnarray}
Notice that this QB
is charged only via an external field instead of a charger, and thus no entanglement exists.
Thus, the diagonal entropy is precisely the coherence.
As shown in Fig. \ref{fig2}(d), the maximally coherent state with coherence $C=2$ is achieved at $t=T/2$, and the coherent  work $W_e^c$ is half of the optimal work $W(\ket{\uparrow\uparrow})=2$.
The shortened time comes from the fact that the optimally coherent  work only requires a balanced distribution of population instead of the battery being the highest energy state.

\emph{Discussion and conclusion}.--
 We have demonstrated the role of the battery-charger entanglement and  the battery's coherence played in extractable work  for general QBs. 
We find that nonvanishing entanglement or coherence during the charging process is necessary for nonzero extractable work.
Instead, at the end of the charging process, entanglement and coherence become obstacles to the final extractable work.
For general QBs, extractable work can be decomposed into  incoherent and coherent parts.
The former is negatively related to not only the battery-charger entanglement but also the battery's coherence, which can be characterized by the diagonal entropy.
The latter is positively related to the coherence in the battery part.
Therefore  coherent  work requires  a 
 nearly balanced population to ensure large overlaps of different energy levels, 
 but incoherent work only requires a population inversion.
 Recent work is devoted to realizing such controls \cite{Mitchison21}.
Our model-independent analysis applies to extractable work, whether it is defined by free energy or ergotropy. 
Our work deepens the understanding of quantum correlations and extractable work in QBs.

\begin{acknowledgments}
This work was supported by the NSFC key Grant No. 12134015, the NSFC (Grants  No. 12275215, No. 11875220,
No. 11874393, No. 12121004, No. 12047502, No. 11975183 and No. 12175178),  the Major Basic Research Program of Natural Science of Shaanxi Province (Grant No. 2021JCW-19), and the Double First-Class University Construction Project of Northwest University.
\end{acknowledgments}

\clearpage

\textbf{Supplemental materials for ``Entanglement, Coherence, and Extractable Work in Quantum Batteries''}
\maketitle

\section{Proof of theorem 1 and 2}
\emph{Theorem 1}.--
For a product pure state $\rho_b(0)\otimes\rho_c(0)$ as the initial state,
if no battery's coherence and no battery-charger entanglement appear during the charging process then extractable work is always zero.

\emph{Proof}.--
Since we always assume that there is no extractable work for the initial state then it is enough to show that the battery's state after charging is the same as the initial state $\rho_b(0)$  if no entanglement and no coherence appear during the charging dynamics.
The condition of vanishing entanglement implies that the charging operator $\mathcal V$ involves no battery-charger interactions.
Furthermore, considering that $\rho_{b}(0)$ is a pure state, we can  write the evolving state of the battery as $\ket{\psi(t)}_b=\sum_n \alpha_n(t)\ket{\varepsilon_n}$ where $\ket{\varepsilon_n}$ are the eigenstates of $\mathcal H_b$.
The vanishing coherence requires that, for a given $t$, only one $\alpha_n(t)=1$ and the rest are 0.
Without loss of generality, we may assume that $\alpha_0(0)=1$ and $\alpha_{n\neq 0}(0)=0$.
To show $\ket{\psi_b(T)}=\ket{\psi_b(0)}$, it is enough to show that $\alpha_0(T)=1$ where $T$ is the charging time.
If this were not the case, then $\alpha_0(T)=0$.
Due to the analyticity of $\alpha_0(t)$
 there must be $0< t_*< T$ such that $0<\alpha_0(t_*)<1$, 
%
 which implies that the battery is coherent at $t_*$.
Thus we arrive at a contradiction.

\emph{Theorem 2}.--
The final extractable work (no matter defined by $W_e(T)$ or $W_f(T)$) is negatively related to the entanglement $S(T)$ for all incoherent QBs.

\emph{Proof}.--
As discussed in the main text, we only consider the ``excellent" QBs whose population of low-energy levels is not greater than that of high-energy levels at the end of the charging process. 
Under this assumption and without loss of generality, we consider two energy level configurations for charged battery states: one is $\rho_b(T)=\sum_{n=0}^N p_n\ket{\varepsilon_n}$ with $p_0\leq p_1\leq \ldots \leq p_N$ and another is $\rho_b'(T)=\sum_{n=0}^Np_n'\ket{\varepsilon_n}\bra{\varepsilon_n}$ with $p_0'=p_0-\delta$, $p_1'=p_1+\delta$, $p_m'=p_m\ (m\geq 2)$ for some $\delta>0$ such that $p_n'\leq p_{n+1}'\ (n\geq 0)$.  
Here  $\ket{\varepsilon_n}$ is the $n$-th energy eigenstate of $\mathcal H_b$, and energy levels $\{\varepsilon_n\}$ are ordered so that $\varepsilon_0<\varepsilon_1<\cdots <\varepsilon_N$.
It is enough to show that  
\begin{eqnarray}
&&\Delta S\equiv S(\rho_b'(T))-S(\rho_b(T))<0,\\ 
&&\Delta W_{e/f}\equiv W_{e/f}(\rho_b'(T))-W_{e/f}(\rho_b(T))>0,
\end{eqnarray}
for completing the proof.
A simple calculation gives
\begin{eqnarray}
&&\Delta S=-(p_0-\delta)\log_2(p_0-\delta)-(p_1+\delta)\log_2(p_1+\delta)\n 
& &+p_0\log_2 p_0+p_1\log_2 p_1<0,\\
&&\Delta W_f=
\delta (\varepsilon_1-\varepsilon_0)-\beta^{-1}\Delta S>0,\\
&&\Delta W_e=\delta (\varepsilon_1-\varepsilon_0)+\delta(\varepsilon_N-\varepsilon_{N-1})>0.
\end{eqnarray}
The first inequality is derived by noticing that $d \Delta S/d\delta <0$ and $\Delta S(\delta=0)=0$.

\end{document}